\documentclass[fleqn,12pt,twoside]{article}
\usepackage{espcrc1}
\usepackage{graphicx}
\usepackage[figuresright]{rotating}
\title{Search for $^{12}$C+$^{12}$C molecule in $^{24}$Mg$^{*}$ populated
by $^{24}$Mg+$^{12}$C} 
\author{C. Beck\address[IReS]{Institut de Recherches Subatomiques, UMR7500,
IN2P3-CNRS/Universit\'e Louis Pasteur, B.P. 28, F-67037 Strasbourg Cedex 2,
France}, A. S\`{a}nchez i Zafra\addressmark[IReS], F. Haas\addressmark[IReS],
P. Papka\addressmark[IReS], V. Rauch\addressmark[IReS], M.
Rousseau\addressmark[IReS], F. Azaiez\addressmark[IReS], P.
Bednarczyk\addressmark[IReS], D. Curien\addressmark[IReS], O.
Dorvaux\addressmark[IReS], A. Nourreddine\addressmark[IReS], J.
Robin\addressmark[IReS], W. von~Oertzen\address[HMI]{Hahn Meitner Institut and
Fachbereich Physik, Freie Universit\"at Berlin, Germany}, B.
Gebauer\addressmark[HMI], Ts. Kokalova\addressmark[HMI], S.
Thummerer\addressmark[HMI], G. de Angelis\address[LNL]{INFN, Laboratori Nationali di
Legnaro, Legnaro and INFN, Padova, Italy}, A. Gadea\addressmark[LNL], S.
Lenzi\addressmark[LNL], D.R. Napoli\addressmark[LNL], S.
Szilner\addressmark[LNL], W.N. Catford\address[Surrey]{University of
Surrey, Guildford, Surrey GU2 5XH, UK}, and D. Jenkins\address[York]{University of York,
Heslington, York YO10 5DD, UK}}
\begin{document}
\maketitle
\begin{abstract}
{\small The $\gamma$-decay properties of $^{24}$Mg excited states are 
investigated in the inverse reaction $^{24}$Mg$+^{12}$C at 
E$_{lab}$($^{24}$Mg) = 130 MeV. At this energy the direct inelastic
scattering populates a $^{24}$Mg$^{*}$ energy region where 
$^{12}$C+$^{12}$C breakup resonances can occur. Very exclusive data 
were collected with the Binary Reaction Spectrometer (BRS) in 
coincidence with {\sc EUROBALL} installed at the {\sc VIVITRON} 
Tandem facility of the IReS at Strasbourg. The experimental detection 
system is described and preliminary results of binary reaction coincidence 
data are presented.}
\end{abstract}
\section{Introduction}
The interpretation of resonant structures observed in the excitation functions
in various combinations of light $\alpha$-cluster nuclei in the energy regime
from the barrier up to regions with excitation energies of 30-50~MeV remains a
subject of contemporary debate. In particular, in collisions between two
$^{12}{\rm C}$ nuclei, these resonances have been interpreted in terms of
nuclear molecules~\cite{Bromley60}. However, in many cases these structures
have been connected to strongly deformed shapes and to the clustering
phenomena, predicted from the $\alpha$-cluster model~\cite{Marsh86},
Hartree-Fock calculations~\cite{Flocard84}, and the Nilsson-Strutinsky
approach~\cite{Leander75}. The question whether these molecular resonances
represent true cluster states in the $^{24}$Mg compound system, or whether they
simply reflect scattering states in the ion-ion potential is still unresolved.
Various decay branches from the highly excited $^{24}$Mg$^*$ nucleus, including
the emission of $\alpha$-particles or heavier fragments, are possibly
available. However, $\gamma$-decays have not been observed so far. Actually, the
$\gamma$-ray branches are predicted to be rather small at these excitation
energies, although some experiments have been
reported~\cite{McGrath81,Metag82,Haas97}, which have searched for these very
small branches expected in the range of $10^{-4}~-~10^{-5}$~fractions of the
total width~\cite{Uegaki98}. The rotational bands built on the knowledge of the
measured spins and excitation energies can be extended to rather small angular
momenta, where finally the $\gamma$-decay becomes a larger part of the total
width. The population of such states in $\alpha$-cluster nuclei, which are
lying below the threshold for fission decays and for other particle decays, is
favored in binary reactions, where at a fixed incident energy the composite
nucleus is formed with an excitation energy range governed by the two-body
reaction kinematics. These states may be coupled to intrinsic states of
$^{24}$Mg$^{*}$ as populated by a breakup process (via resonances) as shown in
Refs. ~\cite{Fulton86,Curtis95,Singer00}. The $^{24}$Mg+$^{12}$C
reaction has been extensively investigated by several measurements of the
$^{12}$C($^{24}$Mg,$^{12}$C$^{12}$C)$^{12}$C breakup
channel~\cite{Fulton86,Curtis95,Singer00}. Sequential breakups are
found to occur from specific states in $^{24}$Mg at excitation energies ranging
from 20 to 35 MeV, which are linked to the ground state and also have an
appreciable overlap with the $^{12}$C+$^{12}$C quasi-molecular configuration.
Several attempts \cite{Curtis95} were made to link the
$^{12}$C+$^{12}$C barrier resonances \cite{Bromley60} with the breakup states.
The underlying reaction mechanism is now fairly well
established~\cite{Singer00} and many of the barrier resonances appear to be
correlated indicating that a common structure may exist in both instances. This
is another indication of the possible link between barrier resonances and
secondary minima in the compound nucleus. 
\section{Experimental set-up of the BRS}
The study of particle-$\gamma$ coincidences in binary reactions in reverse
kinematics is probably a unique tool for the search for extreme shapes related
to clustering. In this way the $^{24}$Mg+$^{12}$C reaction has been
investigated with high selectivity at E$_{lab}$($^{24}$Mg) = 130 MeV with the
Binary Reaction Spectrometer~\cite{Thummerer00} (BRS) in coincidence with {\sc
EUROBALL IV} (EB) installed at the {\sc VIVITRON} Tandem facility of the IReS at
Strasbourg. The choice of the $^{12}{\rm C}(^{24}{\rm Mg},^{12}{\rm
C})^{24}{\rm Mg^{*}}$ reaction implies that for an incident energy of 130~MeV
an excitation energy range up to 30~MeV in $^{24}$Mg is
covered~\cite{Curtis95}. The BRS gives access to a novel approach to the study
of nuclei at large deformations. The excellent channel selection capability of
binary and/or ternary fragments gives a powerful identification among the
reaction channels, implying that EB is used mostly with one or
two-fold multiplicities, for which the total $\gamma$-ray efficiency is very
high. A schematic lay-out of the actual experimental set-up of the BRS with
EB is shown in Fig.~3 of Ref.~\cite{Thummerer00}. The BRS trigger
consists of a kinematical coincidence set-up combining two large-area heavy-ion
telescopes. Both detector telescopes comprise each a two-dimensional position
sensitive low-pressure multiwire chamber in conjunction with a Bragg-curve
ionization chamber. All detection planes are four-fold subdivided in order to
improve the resolution and to increase the counting rate capability (100
k-events/s). The two-body Q-value has been reconstructed using events for which
both fragments are in well selected states chosen for spectroscopy purposes as
well as to determine the reaction mechanism responsible for the population of
these particular states. Fig.~1 displays a typical example of a two-dimensional
Bragg-Peak versus energy spectrum obtained for the $^{24}$Mg+$^{12}$C reaction.
This spectrum shows the excellent charge discrimination achieved with the
Bragg-curve ionization chambers. The Z = 12 gate, which is shown in Fig.~1, will
be used for the processing of the $^{24}$Mg $\gamma$-ray spectra of interest. 
\begin{figure}
  \begin{center}
    \includegraphics[width=0.44\textwidth]{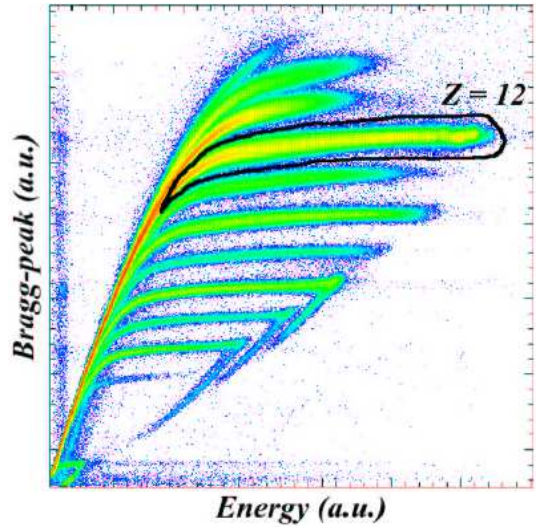}
    \caption{\small\em Two dimensional Bragg-Peak versus energy spectrum, using
                       particle-particle coincidences, measured in
                       $^{24}$Mg(130 MeV)+$^{12}$C reaction
                       with the BRS. The marked Z=12 gate is used to
                       trigger the $^{24}$Mg $\gamma$-ray spectrum
                       displayed in Fig.~2.}
  \end{center}
  \label{Fig.1}
\end{figure}
\section{Preliminary results and subsequent analysis}
The inverse kinematics of the $^{24}$Mg+$^{12}$C reaction and the negative
Q-values give ideal conditions for the trigger on the BRS, because the angular
range is optimum for 12$^\circ$-40$^\circ$ in the lab-system (with the range of
12$^\circ$ to 25$^\circ$ for the recoils) and because the solid angle
transformation gives a factor~10 for the detection of the heavy fragments. Thus
we have been able to cover a large part of the angular distribution of the
binary process with high efficiency, and a selection of events in particular
angular ranges has been achieved. In binary exit-channels the exclusive
detection of both ejectiles allows precise Q-value determination, Z-resolution
and simultaneously optimal Doppler-shift correction. Fig.~2 displays a
Doppler-corrected $\gamma$-ray spectrum for $^{24}$Mg events in coincidence
with the Z=12 gate defined in the Bragg-Peak versus energy spectrum of Fig.~1. All
known transitions of $^{24}$Mg~\cite{Beck01,Wiedenhover01} can be identified in
the energy range depicted. As expected we see decays feeding the yrast line of
$^{24}$Mg up to the 8$^{+}_{2}$ level. The population of some of the observed
states, in particular, the 2$^{+}$, 3$^{+}$ and 4$^{+}$ members of the K=2
rotational band, appear to be selectively enhanced. The strong population of
the K=2 band has also been observed in the $^{12}$C($^{12}$C,$\gamma$)
radiative capture reaction~\cite{Jenkins03}. Furthermore, there is an
indication of a $\gamma$ ray around 5.95 MeV which may be identified with the
10$^{+}_{1}$ $\rightarrow$ 8$^{+}_{2}$ transition as proposed in
Ref.~\cite{Wiedenhover01}. It has been checked in the $\gamma$-$\gamma$
coincidences that most of the transitions shown in Fig.~2 belong to cascades 
which contain
the characteristic 1368 keV $\gamma$ ray and pass through the lowest 2$^{+}$
state in $^{24}$Mg. Still, a number of transitions in the high-energy part of
the spectrum (6~MeV~-~8~MeV) have not been clearly identified. 
The reason why the search for a $\gamma$-decay in $^{12}$C+$^{12}$C has not 
been conclusive so far~\cite{McGrath81,Metag82,Haas97} is due to the excitation
energy in $^{24}$Mg as well as the spin region (8$\hbar$-12$\hbar$) which were
chosen too high. The next step of the analysis will be the use of the BRS
trigger in order to select the excitation energy range by the two-body Q-value
(in the $^{12}$C+$^{24}$Mg channel), and thus we will be able to study the
region around the decay barriers, where
$\gamma$-decay becomes observable. According to recent predictions~\cite{Uegaki98}
$\gamma$-rays from 6$^{+}$ $\rightarrow$ 4$^{+}$ should have measurable
branching ratios. Work is currently in progress to analyse the $\gamma$ rays
from the $^{12}$C($^{24}$Mg,$^{12}$C $^{12}$C)$^{12}$C ternary breakup
reaction. \\ 
\noindent
{\small
{\bf Acknowledgments:} We thank the staff of the {\sc VIVITRON} for providing
us with good $^{24}$Mg stable beams, M.A. Saettel for the targets, and J. Devin
for the excellent support in carrying out the experiment. This work was
supported by the french IN2P3/CNRS, the german ministry of research (BMBF grant
Nr.06-OB-900), and the EC Euroviv contract HPRI-CT-1999-00078.} 
\begin{figure}
  \begin{center}
    \includegraphics[width=0.68\textwidth]{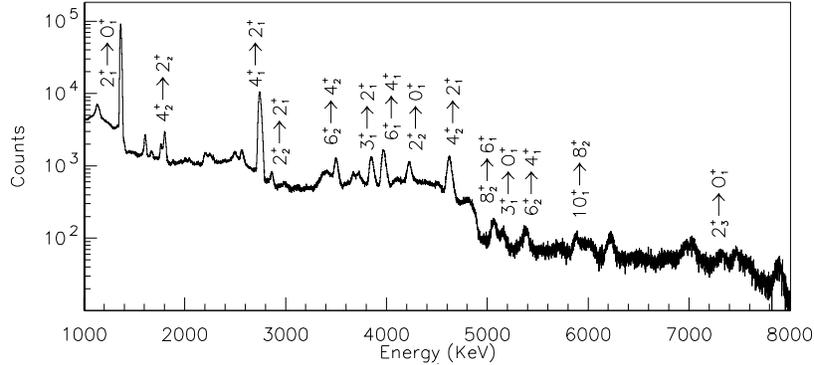}
    \caption{\small\em Doppler corrected $\gamma$-ray spectrum for $^{24}$Mg,
                       using particle-particle-$\gamma$ coincidences, measured
                       in the $^{24}$Mg(130 MeV)+$^{12}$C reaction
                       with the BRS/EB detection system (see text). }
  \end{center}
  \label{Fig.2}
\end{figure}

\end{document}